\documentclass{article}
\def\del{\partial}
\def\ie{{\it i.e.}}

\def\reals{{\mathbb R}}
\newcommand\nxt[1] {\\\raisebox{.12em}{\rule{.35em}{.35em}}\mbox{\hspace{0.6em}}#1}

\usepackage{fortschritte}
\usepackage{amsmath,epsfig,psfrag,amssymb}
\def\beq{\begin{equation}}                     %
\def\eeq{\end{equation}}                       %
\def\bea{\begin{eqnarray}}                     
\def\eea{\end{eqnarray}}                       
\begin {document}                 
\def\titleline{The Tachyon does Matter}
\def\email_speaker{{\tt walcher@kitp.ucsb.edu}}
\def\authors{
%
%
%
%
%
A. Buchel\1ad, J. Walcher\2ad\sp}

\def\addresses{
%
%
%
%
\1ad                                        %
Michigan Center for Theoretical Physics\\
Randall Laboratory of Physics, The University of Michigan \\
Ann Arbor, MI 48109-1120, USA\\
\2ad                                        
Kavli Institute for Theoretical Physics\\
University of California \\
Santa Barbara, CA 93106, USA \\
}
\def\abstracttext{
%
%
We review the concept of S-branes introduced by Gutperle
and Strominger \cite{gust}. Using the effective spacetime
description of the rolling tachyon worldsheets discussed by
Sen, we analyze the possibility that the gravitational backreaction
of tachyon matter is important in the time-dependent
process. We show that this is indeed the case in the example
of the S0-brane in 4-dimensional Einstein-Maxwell theory.
This talk is based on \cite{paper}.
}
\large
\makefront
\section{S-branes}

Gutperle and Strominger \cite{gust} have argued that there
should be spacelike branes in string theory. Let us start
by reviewing what one means by a ``spacelike brane'' (S-brane).

\subsection{Spacelike kinks}

It is well-known that scalar field theories in $1+1$ dimensions,
$$
S = \int -\frac 12(\del_t\Phi)^2 + \frac12(\del_x\Phi)^2 + V(\Phi)\,,
$$
have static solitonic kink solutions that interpolate between the 
minima of the appropriately chosen $V$. For example, in $\Phi^4$
theory, there is a soliton that corresponds to the classical 
trajectory of a particle moving in the inverted potential $-V$
from one extremum to the other.

In string theory, if $\Phi$ is the tachyon on an unstable D$(p+1)$-brane,
Sen has shown \cite{sen} that such a tachyonic kink corresponds to a 
stable D$p$-brane. This is essentially due to a term
\begin{equation}
\int dT\wedge C^{(p+1)}
\label{RRcoup}
\end{equation}
in the effective action of the unstable brane \cite{sen,billo}, where 
$C^{(p+1)}$ is the $p+1$-form RR field coupling to D$p$-brane charge.

Gutperle and Strominger argue in \cite{gust} that one can also make
spacelike kinks, in the following way. Consider spatially homogeneous 
initial conditions for the scalar field at the top of its potential and 
push it infinitesimally towards one minimum. If the scalar field is 
weakly coupled to some form of classical radiation, it will loose energy 
and for $t\to\infty$ eventually settle down in the minimum of the potential. 
The time reversed process looks like (finely tuned) radiation coming in 
from infinity and exciting the tachyon to the top of its potential. The 
total process\footnote{which might seem highly unlikely due to the necessary
fine tuning. One can understand this as a resonance.}, in which the field 
goes from one minimum to the other looks like a scalar field kink {\it 
in time}. 

In string theory, a spacelike tachyonic kink is a source for RR fields,
again because of (\ref{RRcoup}). In the limit that the lifetime of the 
resonance is very short, the object corresponding to it can be thought of 
as a spacelike analog of the usual timelike branes. (One can also think of 
this as a tachyonic brane, since it moves outside of the lightcone). For 
the purposes of this talk, one may use the following as a working definition.%
\footnote{There are also codimension two S-branes obtained from tachyonic 
vortices, but we shall not discuss those here.}

\smallskip

{\it \noindent An S$p$-brane is the time-dependent process involving the
creation and subsequent decay of an unstable D$(p+1)$-brane in string
theory.}

\smallskip

The fact that spacelike branes as spacelike tachyonic kinks require the 
coupling of the scalar to radiation is a crucial difference to the
timelike case that is important to keep in mind.

The main motivation for introducing S-branes comes from the desire to
generalize holography to the time-dependent and, in particular, 
cosmological situation. In string theory, holography is realized as a 
correspondence between open string or brane physics on the one side and 
closed string or bulk physics on the other side. In light of this, a 
spacelike brane on which open strings can end with Dirichlet boundary 
conditions in the time direction, is the natural ``holographic plate'' 
for cosmological, and, in particular, de Sitter spacetimes. This desirable
connection of S-branes to dS/CFT \cite{strominger,witten} gives a 
justification for the R-symmetry requirements imposed on the gravity 
solutions that we discuss below.

\subsection{The S-brane charge}
                                            
It is worthwhile mentioning here that there is nothing funny about 
charges of spacelike objects. Given a spacelike source for a $(p+1)$-form
field (in flat space),
\begin{equation}
dF^{(p+2)}=0\qquad\qquad *\!d\!*F^{(p+2)} = Q \delta(x_\perp) dx_{||} \,,
\label{pform}
\end{equation}
one can measure the charge $Q$ by integrating over a sphere surrounding 
the position of the source,
\begin{equation}
Q=\int_{S^{D-p-2}} * F^{(p+1)} \,.
\label{charge}
\end{equation}
The statement that the charge is conserved refers to the
fact that $Q$ is independent of the sphere used to compute it.
This is independent of whether the source is space- or timelike.
There is one important difference, however. In solving (\ref{pform})
one needs the propagator of a massless scalar in the transverse directions.
From the definition of an S-brane, one should like to use a causal Green's
function (advanced plus retarded, say). Now the causal propagator for a
massless scalar has support on the lightcone for an even number of transverse
dimensions and support inside the lightcone for an odd number of transverse
directions. This is likely to complicate the construction of an S-brane 
with an even number of transverse directions, like the expected S-branes 
in type IIB string theory, substantially. Mentioned in \cite{gust}, this
difference does not seem to have been properly taken into account in the 
literature so far.

\subsection{Gravitational solutions}

\begin{figure}
\begin{center}
\psfrag{infty}{$t=-\infty$}
\psfrag{zero}{$t=0$}
\psfrag{tzero}{$t=Q$}
\epsfig{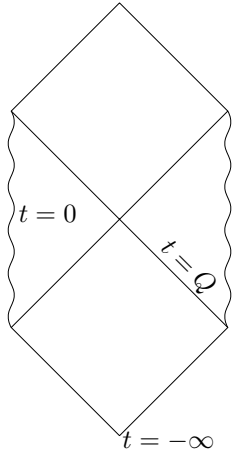}
\caption{Penrose diagram for the metric (\ref{metric}). The wavy
lines are time-like curvature singularities.}
\label{pendg}
\end{center}
\end{figure}

Gravitational solutions corresponding to S-branes have been proposed
in \cite{cgg} and \cite{kmp}, further generalized, for instance, in
\cite{sbrane1,sbrane2,sbrane3}. Some of their features make their 
interpretation as S-branes, in fact, rather difficult. We shall exemplify 
these difficulties for the S$0$-brane in 4-dimensional Einstein-Maxwell 
gravity \cite{gust},
\begin{equation}
S=\int \sqrt{-g} \bigl(R-\frac14F^4\bigr) \,.
\label{em}
\end{equation}
The natural ansatz has SO$(2,1)\times\reals$ symmetry and is of the form
\begin{equation}
ds^2 = -c_1^2dt^2 + c_2^2 dz^2 + c_3^2 ds^2_{H_2}\,,
\label{metric}
\end{equation}
in which the warp factors $c_1$, $c_2$, and $c_3$ are functions of $t$ only, 
and the two-dimensional hyperbolic space $H_2$ gives the leaves of the
transverse foliation. From flat space considerations, one is led to
make the ansatz
\begin{equation}
*F = Q {\it vol}_{H_2}
\label{flux}
\end{equation}
for the flux, where ${\it vol}_{H_2}$ is the volume form on $H_2$. The 
resulting equations of motions become
\begin{eqnarray}
-\frac 1{c_1^2} \biggl[\frac{2c_3''}{c_3} + \frac{c_2''}{c_2}
-\frac{c_1'}{c_1} \Bigl(\frac{2c_3'}{c_3}+\frac{c_2'}{c_2}\Bigr)
\biggr] &=& \frac{Q^2}{c_3^4} 
\label{eqmot1} \\
\frac 1{c_1^2}\biggl[\frac{c_2''}{c_2} - \frac{c_2'}{c_2}\frac{c_1'}{c_1}
+\frac{2c_2'}{c_2}\frac{c_3'}{c_3}\biggr] &=& -\frac{Q^2}{c_3^4} 
\label{eqmot2} \\
-\frac 1{c_3^2} + \frac 1{c_1^2}\biggl[\frac{c_3''}{c_3} 
-\frac{c_3'}{c_3}\frac{c_1'}{c_1} + \Bigl(\frac{c_3'}{c_3}\Bigr)^2
+\frac{c_3'}{c_3}\frac{c_2'}{c_2}\biggr] &=&\frac{Q^2}{c_3^4} \,,
\label{eqmot3}
\end{eqnarray}
and can be solved explicitly in the gauge $c_1=1/c_2$, to yield 
$c_3=t$, and, imposing $t\to -t$ symmetry,
\begin{equation}
c_2^2 = 1-\frac {Q^2}{t^2}  \,.
\label{rn}
\end{equation}
The global structure of this spacetime is illustrated by the Penrose
diagram in Fig.\ \ref{pendg}. It has flat asymptotics at $t\to\infty$,
a ``horizon''-like coordinate singularity at $t=Q$, and {\it timelike}
curvature singularities at $t=0$. In fact, this diagram is nothing but
the $\pi/2$-rotation of the Reissner-Nordstr\"om black hole, and
(\ref{rn}) can be obtained simply by analytical continuation from the
static solution. This is discussed in more detail, and for many other
solutions, in \cite{timedependent}.

There are a number of problems with the S-brane interpretation of
this solution. The most obvious is presumably that the singularities
are timelike, and not spacelike, which would have been the natural
expectation. Moreover, there are no regions that are causally disconnected
from the singularities, another natural expectation from the definition
of S-branes. And where is the brane, anyway? According to (\ref{charge}),
one would detect this by integrating $F$ over large spheres. However, it
is easy to see that the spacetime in Fig.\ \ref{pendg} does not have any
such spheres! 

For a different interpretation of the diagram in Fig.\ \ref{pendg} involving
negative tension branes, see \cite{quevedo1,quevedo2}.

\section{Tachyon Matter}

In perturbative string theory, S-branes are expected to be described by
imposing Dirichlet boundary conditions in time on the open string 
worldsheet. Sen \cite{sen1,sen2,sen3} has clarified that this is indeed
related to the ``rolling tachyon'' picture of S-branes. At the linearized
level, \ie, for early times, the rolling tachyon with mass squared $-1$
looks like  $T(x^0)=\lambda \cosh x^0$, where $x^0$ is time, and 
$\lambda$ is the initial displacement of the tachyon away from the top
of its potential. From the string worldsheet, this looks like a boundary
interaction
$$
\Delta S = \lambda \int d\tau  \cosh X^0(\tau) \,.
$$
Using Wick rotation and results from boundary CFT \cite{resc}, Sen shows 
that, firstly, this boundary interaction is exactly marginal for all 
values of $\lambda$, and secondly, that such a boundary theory produces
a source for closed strings whose energy-momentum tensor is characterized
by constant energy density and exponentially (in time) vanishing pressure.
Moreover, the tachyon energy-momentum tensor is localized on the plane
of the decaying D-brane. This is essentially due to the fact that the
tachyon is an open string.

This behaviour of the tachyon can be reproduced in an effective field
theory description with a DBI type action \cite{garousi,bergshoeff,ghy},
\begin{equation}
S = \int d^{p+2}x V(T) \sqrt{-\det(g_{\mu\nu}+\del_\mu T\del_\nu T)}
\,.
\label{tm}
\end{equation}
Even though the potential to be used in (\ref{tm}) is not known exactly,
it is characterized by the universal asymptotics $V(T)\to e^{-|T|/\sqrt{2}}$
for $|T|\to\infty$. For purely time-dependent configurations, energy 
density and pressure corresponding to this action are given by
\begin{eqnarray}
\rho &=& \frac{V(T)}{\sqrt{1+(T')^2/g_{00}}}
\label{rho} \\
p &=& - V(T) \sqrt{1+ (T')^2/g_{00}} \,.
\label{p}
\end{eqnarray}
If $\rho={\rm const.}$ with $T$ approaching the minimum of $V$ at $t\to\infty$,
then, since $g_{00}=-1$, $T'$ must approach $1$ up to exponentially small
terms. This implies that $p$ vanishes exponentially. These properties 
derived from the action (\ref{tm}) are called ``tachyon matter''.

\section{Can the tachyon matter?}

Let us summarize what we have discussed so far. 
\nxt S-branes as solitonic kinks can only exist if the scalar field
(the tachyon) is coupled to some form of closed string radiation.
\nxt The gravity solutions found so far do not seem to have the
correct global and singularity structure.
\nxt From the open string perspective, the roll of the tachyon (at zero
string coupling) is characterized by exponentially vanishing pressure and 
constant energy density, localized  in the plane of the decaying brane.
\nxt These properties can be reproduced using an effective DBI-type action.
\\\noindent
It is quite natural, then, to ask for the gravitational backreaction
of tachyon matter on the proposed S-brane backgrounds. This is
achieved by coupling gravity, for instance the Einstein-Maxwell Lagrangian
(\ref{em}), to the tachyon matter (\ref{tm}). This was the essential idea 
of the paper \cite{paper}, and we turn to its consequences for the above
S$0$-brane now.

The coupled system of Einstein-Maxwell gravity and $1$-brane has action
\begin{equation}
S = \int\sqrt{-g}\bigl(R-\frac14 F^4\bigr) 
- \int d^4 x \rho(x_\perp)\Bigl[V(T) \sqrt{-\det(g_{\mu\nu}+\del_\mu T\del_\nu T)}
+ f(T) dT\wedge A \Bigr]\,,
\label{coupled}
\end{equation}
where the last term is the coupling between gauge field and tachyon, as in 
(\ref{RRcoup}). The functions $V$ and $f$ are assumed to have the universal 
exponentially vanishing asymptotics discussed above. Moreover, $\rho(x_\perp)$
is the density of $1$-branes which are smeared over the transverse space.
This smearing, with $\rho(x_\perp)\propto \sqrt{g_\perp}$, is necessary to make 
the equations tractable. The essential change in the equations of motion
is the addition of terms like $(\rho\pm p)/c_3^2$ to the right hand side of 
(\ref{eqmot1}-\ref{eqmot3}). The Maxwell and tachyon equation read
\begin{eqnarray}
(c_3^2 A)' &=& f(T) T' \\
\rho' &=& - (\rho+p) \frac {c_2'}{c_2} - f(T) T' A \,,
\end{eqnarray}
where $\rho$ and $p$ are given by (\ref{rho}) and (\ref{p}).

What are the changes to the solution? First of all, it is easy to 
convince oneself that the early/late time asymptotics are unchanged, 
up to logarithmic corrections. But something interesting happens close
to the coordinate singularity, $t=Q$ in (\ref{rn}).

To see this, let us assume that the warp factors vanish at $t=t_*$
according to $c_2^2 = c_1^{-2}= (t-t_*) \cdot(\text{finite at 
$t=t_*$})$, with $c_3^2$ finite at $t=t_*$. Locally, this looks
like the ``Milne universe'', recently popular due to its
importance in ekpyrotic cosmology. More precisely, introducing
$\tau=\sqrt{t-t_*}$, the metric looks like $-d\tau^2+\tau^2 dz^2$,
which is isomorphic to a piece of two-dimensional Minkowski space.

{} From the expressions for $\rho$ and $p$, (\ref{rho}) and (\ref{p}),
it follows that $\rho=-p$ near $t=t_*$, {\it if $T'$ is finite there}.
The resulting equations (in which the tachyon is essentially
replaced by a cosmological constant), can be solved explicitly.
This finiteness of $T'$, however, is {\it non-generic}. To see this,
we consider the tachyon equation of motion
\begin{equation}
\rho' = -(\rho+p) \frac{c_2'}{c_2} \,,
\end{equation}
and set $\rho+p=\rho (T'/c_1)^2$, which is justified if $T$ itself 
is finite near $t_*$. As a consequence,
\begin{equation}
\rho' = -\frac{1}{2(t-t_*)} \bigl(\rho-\frac 1\rho\bigr) \,,
\end{equation}
which implies $\rho^2 \propto C/(t-t_*) + 1$. This
diverges as $t\to t_*$, unless $C$ is fine tuned to vanish.
Thus the stress tensor of the tachyon diverges, and one expects
a curvature singularity. Indeed, one can solve for the warp
factors, and finds
\begin{equation}
c_1 c_2 \sim 1+ {\rm const.} \sqrt{t-t_*}
\qquad\quad
c_3\sim 1+ {\rm const.} \sqrt{t-t_*} \,,
\end{equation}
leading to a spacelike curvature singularity.

Lastly, one can show that the remaining singularity structure is 
not modified by inclusion of tachyon matter \cite{paper}. In particular, 
the timelike curvature singularity is not resolved in our model. Now, 
however, there is a curvature singularity associated with the tachyon 
{\it before} one reaches the singularity associated with the gauge field, 
so that the order of the latter has shifted.

\section{Conclusions and further developments}

Summarizing, we have seen that the tachyon matter produces non-negligible
backreaction on the S-brane background. While the resulting spacetimes
may still not look exactly like what one would expect, the changes
in singularity structure induced by the tachyon go in the right direction.
There are a few obvious generalizations that one might want to explore,
including D$\overline {\rm D}$-brane systems, to see the effect of a
complex tachyon, or higher dimensional systems, to see the effect of
including a dilaton.\footnote{We have been informed by F.\ Leblond
that the behaviour in higher dimensions can be significantly different 
upon inclusion of the dilaton \cite{amanda}.}

We conclude the talk with a brief overview of some further developments 
concerning S-branes and tachyon matter.

Tachyon matter has been explored from the point of view of boundary
string field theory \cite{terashima1,terashima2}. It has also been 
analyzed in toy models of open string field theory in \cite{zwiebach},
focusing on the problems associated with infinitely many time
derivatives. In \cite{time}, see also \cite{sen4}, it was proposed
to use the tachyon for an emergent definition of time in the context
of canonical quantization of gravity. In particular, it was shown that all
solutions of tachyon matter are at late times equivalent to configurations
of non-interacting non-rotating dust. In anticipation of dS/CFT, S-brane
worldvolume actions have been proposed in \cite{action}.

A more puzzling aspect of the decay process of D-branes and the roll of
the tachyon has been revealed in recent studies of radiation production
rates. While we have here studied the simplest possible coupling of the 
tachyon to closed strings, including only the massless supergravity modes, 
it is a valid question to ask whether this is a good approximation at all. 
In \cite{mukho,okuda} it was argued that {\it massive} closed string modes
may in fact play a crucial role in the decay of the D-brane, as it was shown
that the coupling of the tachyon matter to these modes grows exponentially in
time and with closed string mode level. In \cite{open}, adopting the open
string perspective, the quantum open string production in the exact CFT backgrounds 
of \cite{sen1,sen2,sen3} was computed and shown to diverge due to the exponentially
growing density of open string states. Quantum effects being closed strings,
this again points to the importance of closed strings, see also \cite{gravity}.
These results are disturbing as they indicate that there are fundamental
difficulties in finding a valid approximation scheme.

Lastly, we should also mention the considerable cosmological interest
that tachyon matter has attracted, which, however, we have no space to
review here. See \cite{cosmology} for early literature on tachyon matter
cosmology.

{\bf Acknowledgement} We would like to thank Peter Langfelder
for collaboration on the results of \cite{paper}. J.W.\ would like to
thank the organizers of the 35th Symposium Ahrenshoop for a very
stimulating atmosphere at the meeting, and the Perimeter
$\hat{\,\text{\raisebox{.66em}{\rule{.8em}{.04em}}}\!\!\!\!\!\!{\rm P}{\rm I}}$
Institute for hospitality while these notes were being written up. This work
was supported in part by the NSF under Grant Nos.\ PHY00-98395 (A.B.) and
PHY99-07949 (A.B.\ and J.W.).


\begin{thebibliography}{77}


\bibitem{gust}
M.~Gutperle and A.~Strominger,
``Spacelike branes,''
JHEP {\bf 0204}, 018 (2002)
[arXiv:hep-th/0202210].

\bibitem{sen1}
A.~Sen
``Rolling tachyon,''
JHEP {\bf 0204}, 048 (2002)
[arXiv:hep-th/0203211].

\bibitem{sen2}
A.~Sen
``Tachyon matter,''
arXiv:hep-th/0203265.

\bibitem{sen3}
A.~Sen
``Field theory of tachyon matter,''
arXiv:hep-th/0204143.

\bibitem{resc}
A.~Recknagel and V.~Schomerus,
``Boundary deformation theory and moduli spaces of D-branes,''
Nucl.\ Phys.\ B {\bf 545}, 233 (1999)
[arXiv:hep-th/9811237].

\bibitem{cgg}
C.~M.~Chen, D.~V.~Gal'tsov and M.~Gutperle,
``S-brane solutions in supergravity theories,''
arXiv:hep-th/0204071.

\bibitem{kmp}
M.~Kruczenski, R.~C.~Myers and A.~W.~Peet,
``Supergravity S-branes,''
JHEP {\bf 0205}, 039 (2002)
[arXiv:hep-th/0204144].

\bibitem{strominger}
A.~Strominger,
``The dS/CFT correspondence,''
JHEP {\bf 0110}, 034 (2001)
[arXiv:hep-th/0106113].

\bibitem{witten}
E.~Witten,
``Quantum gravity in de Sitter space,''
arXiv:hep-th/0106109.

\bibitem{sbrane1}
S.~Roy,
``On supergravity solutions of space-like Dp-branes,''
JHEP {\bf 0208}, 025 (2002)
[arXiv:hep-th/0205198].

\bibitem{sbrane2}
N.~S.~Deger and A.~Kaya,
``Intersecting S-brane solutions of D = 11 supergravity,''
JHEP {\bf 0207}, 038 (2002)
[arXiv:hep-th/0206057].

\bibitem{sbrane3}
K.~Ohta and T.~Yokono,
``Gravitational approach to tachyon matter,''
arXiv:hep-th/0207004.

\bibitem{sen}
A.~Sen,
``Non-BPS states and branes in string theory,''
arXiv:hep-th/9904207.

\bibitem{billo}
M.~Billo, B.~Craps and F.~Roose,
``Ramond-Ramond couplings of non-BPS D-branes,''
JHEP {\bf 9906}, 033 (1999)
[arXiv:hep-th/9905157].

\bibitem{garousi}
M.~R.~Garousi,
``Tachyon couplings on non-BPS D-branes and Dirac-Born-Infeld action,''
Nucl.\ Phys.\ B {\bf 584}, 284 (2000)
[arXiv:hep-th/0003122].

\bibitem{bergshoeff}
E.~A.~Bergshoeff, M.~de Roo, T.~C.~de Wit, E.~Eyras and S.~Panda,
``T-duality and actions for non-BPS D-branes,''
JHEP {\bf 0005}, 009 (2000)
[arXiv:hep-th/0003221].

\bibitem{ghy}
G.~W.~Gibbons, K.~Hori and P.~Yi,
``String fluid from unstable D-branes,''
Nucl.\ Phys.\ B {\bf 596}, 136 (2001)
[arXiv:hep-th/0009061].

\bibitem{terashima1}
T.~Takayanagi, S.~Terashima and T.~Uesugi,
``Brane-antibrane action from boundary string field theory,''
JHEP {\bf 0103}, 019 (2001)
[arXiv:hep-th/0012210].

\bibitem{terashima2}
S.~Sugimoto and S.~Terashima,
``Tachyon matter in boundary string field theory,''
JHEP {\bf 0207}, 025 (2002)
[arXiv:hep-th/0205085].

\bibitem{timedependent}
A.~Buchel, P.~Langfelder and J.~Walcher,
``On time-dependent backgrounds in supergravity and string theory,''
Phys.\ Rev.\ D, in press
[arXiv:hep-th/0207214].

\bibitem{paper}
A.~Buchel, P.~Langfelder and J.~Walcher,
``Does the tachyon matter?,''
Annals Phys.\  {\bf 302}, 78 (2002)
[arXiv:hep-th/0207235].

\bibitem{zwiebach}
N.~Moeller and B.~Zwiebach,
``Dynamics with infinitely many time derivatives and rolling tachyons,''
arXiv:hep-th/0207107.

\bibitem{sen4}
A.~Sen,
``Time evolution in open string theory,''
arXiv:hep-th/0207105.

\bibitem{amanda}
F.~Leblond and A.~Peet,
to appear

\bibitem{open}
A.~Strominger,
``Open string creation by S-branes,''
arXiv:hep-th/0209090.

\bibitem{mukho}
P.~Mukhopadhyay and A.~Sen,
``Decay of unstable D-branes with electric field,''
JHEP {\bf 0211}, 047 (2002)
[arXiv:hep-th/0208142].

\bibitem{okuda}
T.~Okuda and S.~Sugimoto,
``Coupling of rolling tachyon to closed strings,''
Nucl.\ Phys.\ B {\bf 647}, 101 (2002)
[arXiv:hep-th/0208196].

\bibitem{time}
A.~Sen,
``Time and tachyon,''
arXiv:hep-th/0209122.

\bibitem{gravity}
B.~Chen, M.~Li and F.~L.~Lin,
``Gravitational radiation of rolling tachyon,''
arXiv:hep-th/0209222.

\bibitem{action}
K.~Hashimoto, P.~M.~Ho and J.~E.~Wang,
``S-brane actions,''
arXiv:hep-th/0211090.

\bibitem{cosmology}
G.~W.~Gibbons,
``Cosmological evolution of the rolling tachyon,''
Phys.\ Lett.\ B {\bf 537}, 1 (2002)
[arXiv:hep-th/0204008]; \ \
%
M.~Fairbairn and M.~H.~Tytgat,
``Inflation from a tachyon fluid?,''
arXiv:hep-th/0204070;
%
S.~Mukohyama,
``Brane cosmology driven by the rolling tachyon,''
Phys.\ Rev.\ D {\bf 66}, 024009 (2002)
[arXiv:hep-th/0204084];
%
A.Feinstein,
``Power-law inflation from the rolling tachyon''
arXiv:hep-th/0204140;
%
T.~Padmanabhan,
``Accelerated expansion of the universe driven by tachyonic matter,''
Phys.\ Rev.\ D {\bf 66}, 021301 (2002)
[arXiv:hep-th/0204150];
%
G.~Shiu and I.~Wasserman,
``Cosmological constraints on tachyon matter,''
Phys.\ Lett.\ B {\bf 541}, 6 (2002)
[arXiv:hep-th/0205003];
%
G.~Shiu, S.~H.~Tye and I.~Wasserman,
``Rolling tachyon in brane world cosmology from superstring field theory,''
arXiv:hep-th/0207119.

\bibitem{quevedo1}
C.~Grojean, F.~Quevedo, G.~Tasinato and I.~Zavala C.,
``Branes on charged dilatonic backgrounds: Self-tuning, Lorentz  violations and cosmology,''
JHEP {\bf 0108}, 005 (2001)
[arXiv:hep-th/0106120].

\bibitem{quevedo2}
C.~P.~Burgess, F.~Quevedo, S.~J.~Rey, G.~Tasinato and C.~.~Zavala,
``Cosmological spacetimes from negative tension brane backgrounds,''
arXiv:hep-th/0207104.


\end{thebibliography}
\end{document}